\documentstyle[12pt,aps,epsf,preprint,tighten]{revtex}

\begin{document}
\draft
\title{
\begin{flushright}
{\small Preprint CNS-9710\\
May 1997}\\
\end{flushright}
Spatiotemporal dynamics of discrete sine-Gordon lattices with
sinusoidal couplings}
\author{
Zhigang Zheng$^{a,b}$\footnote{E-mail address: zgzheng@hkbu.edu.hk.} Bambi Hu
$^{a,c}$ and Gang Hu$^{d,b}$}
\address{
$^a$Department of Physics and Center for Nonlinear Studies, Hong Kong
Baptist University, Hong Kong \\
$^b$Department of Physics, Beijing Normal University, Beijing 100875,
China \\
$^c$Department of Physics, University of Houston, Houston TX77204, USA \\
$^d$Center of Theoretical Physics, Chinese Center of Advanced Science
and Technology (World Laboratory), Beijing 8730, China}
%\date{\today}
\maketitle
 
\begin{abstract}
The spatiotemporal dynamics of a damped sine-Gordon chain with sinusoidal
nearest-neighbor couplings driven by a constant uniform force are discussed.
The velocity characteristics of the chain versus the external force is
shown. Dynamics in the high- and low-velocity regimes are investigated. It
is found that in the high-velocity regime, the dynamics is dominated by
rotating modes, the velocity shows a branching bifurcation feature, while in
the low-velocity regime, the velocity exhibits step-like dynamical
transitions, broken by the destruction of strong resonances.
\end{abstract} 
\pacs{PACS numbers: 05.45.+b, 05.40.+j, 03.20.+i} 
 
%\begin{multicols}{2}
\section{Introduction}
 
Spatiotemporal dynamics in systems with many degrees of freedom gained great
interest during the last twenty years because of their complicated
spatiotemporal patterns and possible applications in many fields, such as
turbulence, neural networks, biology, secure telecommunications,
spatiotemporal control of chaos, stochastic resonances, Josephson-junction
lattices, etc.[1-5]. Researches on the discrete sine-Gordon lattice have
also been a surge of interest recently since it can be used to model many
physical systems, such as dislocations, magnetic and ferromagnetic domain
walls, spin- and charge-density waves, and arrays of Josephson
junctions[6,7]. The discrete sine-Gordon chain corresponds to the so-called
Frenkel-Kontorova (FK) model in the conservative case, which was mainly
studied in exploring the commensurate-incommensurate phase transitions of
the ground state[8] and discussing its dc and ac responses[9,10]. Recently,
the damped dynamics of this model with harmonic coupling were numerically,
theoretically and experimentally investigated in relating to the fluxon
dynamics of one-dimensional Josephson-junction arrays[7,11]. In the damped
case, numerous metastable states exist and play a significant role, and this
will lead to complicated spatiotemporal patterns and dynamics.
 
The damped FK chain driven by an external force can be described by the
equation of motion of a coupled chain of pendula 
\begin{equation}
\begin {array}{ll}
\label{1}&\stackrel{..}{x}_j+\gamma \stackrel{.}{x}_j+\sin x_j=K[V^{\prime
}(x_{j+1}-x_j-a)\\&-V^{\prime }(x_j-x_{j-1}-a)]+F,
\end {array}
\end{equation}
where $V(x)$ describes the coupling law between nearest-neighbor elements, $
\gamma $, $K$, $a$ and $F$ are the friction coefficient, coupling strength,
static length of the chain and the external driving force, respectively. The
coupling mechanism of the discrete sine-Gordon chain is generally nonlinear
(anharmonic) in the applications to real physical systems. For a real chain
of atoms or molecules, the coupling law may be the Lennard-Jones, Toda or
Morse types[12]. For a DNA chain, the couplings of the helices and base
pairs are complicated rotator interactions. In the discussions of
antiferroelectric liquid crystals, the coupling mechanism of dipoles and
layers is also the sinusoidal form[13]. The non-convex coupling cases are
more interesting than the convex cases. In this paper, we concern with the
sinusoidal (rotator) coupling case 
\begin{equation}
\label{2}V(x)=1-\cos x.
\end{equation}
This case is especially realistic when the local degree of freedom is an
angle, like in dipolar or magnetic couplings, where $V(x)$ is a periodic
function. The sinusoidal coupling is one of the simplest forms. This kind of
couplings have been considered for coupled rotator systems[14], base pair
rotations in DNA[13], magnetic Heisenberg models[15] and 1D chiral XY
model[16], granular superconductors[17], and Josephson-junction array
ladders[18]. Moreover, for strong-coupling strengths, the sinusoidal
interaction can be approximated by the harmonic form. In Fig.1, two kinds of
couplings are shown. Only in the vicinity of the equilibrium position will
the two kinds of couplings coincide. It is very interesting to note that
these two cases reduce to the same sine-Gordon equation in the continuum
limit. In fact, there is not a unique way to discretize a continuum
equation. Sometimes nonlinear approaches are more significant because
nonlinear localized modes are intrinsically discrete. Therefore it is
necessary to study nonlinear coupling cases as they may correspond to some
real physically discrete systems.
 
This paper is organized as follows. In Sec. II, we discuss the dynamics in
the high-velocity regime. We observe that in this regime, rotating modes
dominate, as theoretically predicted[19, 20]. The rotating modes result in a
branching bifurcation of the $v-F$ (where $v$ is the average velocity of the 
chain) relation. In Sec. III, dynamics in the
low-velocity regime is analyzed. Dynamical transitions between resonant
steps are investigated, and these steps can be theoretically predicted by
using the harmonic-coupling formula proposed in [11]. We also find that some
steps disappear due to the destruction of travelling-wave solutions. This
leads to the emergence of new dynamical phases. Sec.IV gives some concluding
remarks.
 
\section{High-velocity regime: rotating modes}
 
By inserting $V(x)=1-\cos x$ into (2), we rewrite the equation of motion: 
\begin{equation}
\begin{array}{ll}
\label{3}&\stackrel{..}{x}_j+\gamma \stackrel{.}{x}_j+\sin x_j=K[\sin
(x_{j+1}-x_j-a)\\&-\sin (x_j-x_{j-1}-a)]+F.
\end{array} 
\end{equation}
This equation is highly nonlinear. Only in some limit cases, for example $
K\rightarrow \infty $, can it be analytically treated. Hence we study its
dynamics mainly by using numerical simulations. The fourth-order Runge-Kutta
integration algorithm is used and the time step is adjusted according to the
numerical accuracy. Periodic boundary conditions are added, i.e., $
x_{j+N}(t)=x_j(t)+2\pi M,$ where $M$ is an integer that counts the net
number of kinks trapped in the ring. Therefore the frustration $\delta
=\frac MN$ and the spring constant $a=2\pi \delta $.
 
In the case of large coupling $K$, the system can be well described by the
continuum sine-Gordon chain. In this case, it was found that there exists a
critical chain velocity $v_c=2\pi \delta \sqrt{K}$ that separates two kinds
of dynamics (kinks)[21, 22]. When $v<v_c$, the motion is that of localized
solitons, which is called the {\em low-velocity regime.} When $v>v_c$, the
motion is characterized by a whirling wave, i.e., the moving kink is
strongly extended. We call this region the {\em high-velocity regime}. Let
us first study the dynamics in the high-velocity regime. Fig.2 gives a
typical evolution of the velocities for an eight-particle chain. It has been
found that although there are couplings between the elements, their motions
are inhomogeneous, i.e., some particles rotate with a finite velocity while
others remain pinned in the potential wells. This is a consequence of both
non-convex coupling and bistability. Recalling the phase-space structure of
a single pendulum in the underdamped case, one finds that there are two
kinds of attractors with one fixed points and the other running solutions.
This leads to bistability[23]. However, distinct difference between the
single-particle and sinusoidal-coupling cases can be found in Fig.2. It is
shown that oscillation occurs for both rotating and pinned particles, which
is the result of coupling and this will not happen for the single-particle
case. Our numerical picture also supports the prediction of Takeno and
Peyrard[19], who proved the existence of rotating mode in conservative
sinusoidal-coupled sine-Gordon chain. This was also related to what is 
discussed in [20]. The existence of rotating modes is a consequence of the
particular topology of the sinusoidal coupling, which is also a localized
excitation like breathers.
 
In Fig.3, we plot the average velocity $v$ versus the external driving force 
$F$ for $N=8$, $K=1$ and $M=$1, 2, 3. It can be seen that in the
high-velocity regime, there is a bifurcation of $v$ branches, where each
branch can be expressed by 
\begin{equation}
\label{4}v_i(t)=v_i^0+\delta v_i\sin (\omega t+\varphi _i), 
\end{equation}
where $v_i^0$ describes the rotating frequency of the $i$th particle, and
the second term corresponds to the oscillation around the central frequency, 
$\delta v_i$, $\omega $ and $\varphi _i$ are the oscillation amplitude,
the oscillating frequency and the phase, respectively. The rotating frequency 
can be expressed by 
\begin{equation}
\label{5}v_i^0=\left\{ 
\begin{array}{c}
0 
\text{, for a pinned particle} \\ \frac F\gamma ,\text{ for a rotating
particle.} 
\end{array}
\right. 
\end{equation}
Thus the average velocity (averaged over time and lattice) will be a
quantized one: 
\begin{equation}
\label{6}v=\frac{nF}{N\gamma }, 
\end{equation}
which corresponds to the numerical branches, where $n=$1, ..., $N$. Each 
numerical point is obtained by starting from randomly chosen initial motions 
of the chain. The dotted lines are theoretical results (6). One can find that 
they agree
precisely with the numerical results. The mechanism behind this quantization 
is due to the non-convexity of the coupling. This leads to many metastable
states. It can also be observed from Fig.3
that for $M>1$, the bottom few lines disappear. This is a consequence of
multikinks. The threshold forces for emergences of different lines are not
the same. Lines with larger slopes have larger threshold values. For the $
n=1 $ branch, the threshold $F_c$ is just the threshold of the emergence of
bistability for a single driven damped pendulum. This is also the smallest
threshold.
 
The rotating modes are stable only for the weak coupling case. If one
increases the coupling strength $K$, the dynamics approaches that of a
convex-coupling case. In Fig.4, we give the $v-F$ characteristics for $N=8$, 
$M=1$ and $K=$5, 10. It can be found that branches of $n<N$ disappear, only
the $n=N$ branch survives, i.e., all particles in the chain will rotate.
This is natural when one uses a larger coupling strength, because stronger
couplings will cause the rotating modes to become unstable. The convex part
will strongly affect the rotating modes. It should also be noted that
several unstable regions can be observed on the high-velocity line. These
regions become smaller when one increases $F$. This is a residual effect of
sinusoidal coupling, where non-convex effect can still play a role. In
Fig.5, evolutions of velocities for one of the particles in these regions
are shown. We find that motions in small-$F$ unstable regions are irregular
while they are complicated periodic or quasiperiodic motions in larger-$F$
unstable regions.
 
One may also observe the dynamics in the low-velocity regime in the above
figures. They are more interesting. Hence we now turn discussions to
dynamics in this region.
 
\section{Low-velocity regime: resonant-step dynamical transitions}
 
For a sine-Gordon chain with harmonic coupling, dynamics in the low-velocity
regime exhibit soliton behavior. But because of the discreteness of the
chain, the attractor in this region is a distorted travelling wave, which is
composed of a moving kink and small superimposed oscillating wave. This will
cause the radiation of small linear wave of a moving kink in its wake. We
have given a mean-field description of soliton dynamics in this regime and
theoretically predicted the resonance behavior[11].
 
In the sinusoidal coupling case, similar behaviors only occur for a system
with multikinks ($M>1$). In Fig.3 for $M=1$, we find that in the
low-velocity regime, the velocity remains zero, i.e., the chain remains
pinned until the external force exceeds a critical value that indicates the
emergence of bistability. When one increases the coupling strength $K$, we
find distinct differenences, i.e., steplike resonances occur (see Fig.4, $
K=5 $, 10), which is similar to the harmonic case. In fact, the convex part
of the coupling takes effects in this case. On the other hand, non-convex
effects can still play a role in some regimes. For example, in Fig.4(a),
near the boundary between the high- and low-velocity regimes, non-convex
coupling effects are significant, the zero velocity branch reappears. Also
in Fig.4(b)(inset), the enlarged plot of a step shows a negative slope and a
small unstable gap can be observed. Fig.3(b) and 3(c) show the situations
for $M>1$. Dynamics in the low-velocity regime is rather complicated. In
Fig.3(b), for $M=2$, the depinning force $F_c$ is very small, and zero
velocity reappears around 0.1 to 0.2. This is also a non-convex effect.
 
In Fig.6, we enlarge the low-velocity regime of Fig.3(c) for $M=3$ in order
to give a more precise analysis. In order to make a good comparison, we also
plot the $v-F$ curve for the harmonic case where all parameters remain the
same as the sinusoidal case. For both cases, steplike dynamical transitions
can be observed. Transitions between these states lead to the gaps in Fig.6.
We studied the mechanisms of the gap behavior for the harmonic case. The
existence of resonance steps is a time-scale competition between kink
propagation and its radiated phonon waves. When the two frequencies satisfy
the resonance condition, the mode is locked, hence steps occur. We have
derived a formula for all the steps[11]: 
\begin{equation}
\label{7}v(m_1,m_2)=\frac{m_2}{m_1}\sqrt{\beta +4K\sin {}^2(\frac{m_1\delta
\pi }{m_2})}, 
\end{equation}
where $(m_1,m_2)$ is a pair of integers that describes the resonance between
kinks and linear waves, $\delta $ is the frustration, and $\beta $ is a
contraction factor that we introduced in terms of mean-field treatment to
consider the commensurability effect. Physically $\beta $ can be
interpretted as the depinning force that is needed to overcome the
Pierels-Nabarro (PN) barrier and move continuously the static kink along the
chain[11]. Steps in Fig.6 can be well recognized by using (7) and all
resonance steps are labeled for the harmonic case. It may be found that for
a small force, steps for the sinusoidal case can exist and agree well with
those for the harmonic case. However, there are several regions where the
velocity jump down to a lower branch. This kind of dynamical transitions did
not occur for the harmonic coupling case, hence this is also a typical
non-convex effect. A careful comparison between these two cases indicates
that {\em these drop-offs correspond to the destruction of strong
resonances. }The most significant step that is completely destroyed is the $
1:1$ resonance, where the travelling wave solution becomes unstable. In
addition, $2:1$, $3:2$ and other giant resonance steps are also partly
destroyed. In Fig.7, we give the evolution of $x_i(t)$ in the destructed
regions. A typically good travelling wave is also shown in order to make a
comparison. It is vividly observed that in the resonance-destructed regions,
the travelling wave becomes unstable and disordered motion takes place. This
can be qualitatively understood. {\em While the kink frequency and its
linear wave satisfy the resonance condition, for those strong enough
resonances, they are so strong that the linear wave around the kink becomes
unstable and is amplified, this will in turn destroy the travelling wave,
and eventually the dynamics becomes chaotic and resonant steps disappear.}
The 1:1-destructed region is a typical chaotic region, and this also leads
to the connection between low- and high-velocity regimes. We call the 1:1
branch a {\em disordered phase}. Numerical studies indicate that the motion
on this whole branch is chaotic, and the preserved order is completely
destroyed.
 
It is very interesting to note that this mechanism is valid especially for
incommensurate cases. For example, when $\delta $ is the golden mean $\delta
_G=\frac{\sqrt{5}-1}2$, the $1:1$ resonance will always be destroyed for
intermediate coupling cases, i.e., the disordered phase always exists. This
is another route for the transition between low- and high-velocity regimes,
which occurs only for non-convex cases. In Fig. 8, we give the $v-F$
characteristics of the golden mean approach by the Fibonacci sequence $
\delta =\frac 58$, $\frac 8{13}$, and $\frac{13}{21}$, ..., it can be easily
found that at $F_c\simeq 0.318$, the chain velocity always drops to the
lower disorder phase, where the 1:1 resonance causes the travelling wave to
be unstable.
 
\section{Concluding remarks}
 
In this paper, we have studied the dynamics of the discrete sine-Gordon
chain with a sinusoidal coupling. Complicated spatiotemporal patterns exist
in this nonlinear lattice system. In the high-velocity regime, the
attractors are the rotating waves. The motion of the chain may be
inhomogeneous, i.e., some particles can remain pinned while others rotate.
This is a consequence of both non-convex discreteness and bistability. This
inhomogeneity leads to a branching bifurcation of the $v-F$ characteristics.
For stronger couplings, branching bifurcation is destroyed and only the
collective motion branch survive. Non-convex effect still can be observed
along this branch. In the low-velocity regime, the attractor in a large
region is the travelling wave. This is similar to the harmonic-coupling
case, where the convexity of the interaction leads to travelling wave
propagation. Resonant-step transitions can be well predicted by the formula
(7). However, due to the intrinsic non-convexity of the sinusoidal coupling,
some resonant steps are completely or partly destroyed, leading to the
disordered phase. This causes the travelling wave to be unstable, hence the
motion becomes irregular and even chaotic.
 
The present model and results may be applied to some experimental fields,
such as granular superconductors and Josephson junction ladders. In the
experimental studies of the discrete Josephson transmission lines with
stacked junctions[24] and coupled long Josephson junctions[25], similar
behaviors also exist. Moreover, oscillating breathers and rotating states
have also been found. In the investigation of Josephson junction
ladders[26], the same coupling model corresponding to the 1D chiral XY model
has been proposed [16]. The ground state and relaxation phenomena had been
fully explored. But to our knowledge, no discussions on the damped driven
dynamics of this model have been found. In a real experimental environment,
such considerations may give more insight into the intrinsic properties of
the system. The present work may be a first step in exploring the
spatiotemporal behaviors in this model. Moreover, much knowledge can be
captured in the investigations of this kind of disorder systems, compared to
the continuum sine-Gordon system. An important issue concerns the noise
effect in this system, for fluctuations will induce transitions between
different dynamical states. This problem now is under exploration.
 
\bigskip
\bigskip
 
One of the authors(Zheng) thanks Professors David Stroud and N.M.Plakida for
useful discussions on superconductivities and Josephson-junction arrays and
Prof. L.H.Tang for discussions on ground state problem. He also thanks all
the colleagues in Center for Nonlinear Studies of Hong Kong Baptist
University. This work is supported in part by the Research Grant Council
RGC and the Hong Kong Baptist University Faculty Research Grant
FRG.

\begin{figure}
%\narrowtext
\caption{A schematic plot of two kinds of coupling forms, $V(x)=\frac 12x^2$
and $V(x)=1-\cos (x)$. Only in a small region near the origin, the two
functions coincide.} 
\end{figure}
 
\begin{figure}
\narrowtext
\caption{The evolution of the velocities of an eight-particle system with the
sinusoidal coupling. Rotating modes($v>0$) emerge due to the non-convexity
of the coupling. Both the rotating and pinned particles are oscillatory.} 
\end{figure}
 
\begin{figure}
%\narrowtext
\caption{The $v-F$ characteristics of the sinusoidal coupling cases for $N=8$, 
$\gamma =0.1$, $K=1.0$ and $M=$1, 2, 3. Branching bifurcations are shown in
the high-velocity regime. Resonance-step transitions can also be observed in
the low-velocity regime.} 
\end{figure}
 
\begin{figure}
%\narrowtext
\caption{The $v-F$ plot for $M=1$, $K=5.0$ and $10.0$. Other parameters are
the same as Fig.3. Rotating modes are destroyed due to the role of convex
part of the coupling. Several unstable regions resulting from the non-convex
effect are labeledby boxes.} 
\end{figure}
 
\begin{figure}
%\narrowtext
\caption{The motion of one particle in the unstable boxes of Fig.4(a).
Quasiperiodic motion can be observed for smaller forces. In the unstable
regions with larger forces, the motion is modulated periodic.}
\end{figure}
 
\begin{figure}
%\narrowtext
\caption{The $v-F$ relations for both harmonic (circles) and sinusoidal
(crosses) cases with the same parameters: $N=8$, $M=3$, $K=1.0$ and $\gamma $
$=0.1$. Step transitions can be observed, and all steps for the
harmonic-coupling case are labeled by resonances $(m_1,m_2)$. Good agreement
is shown for small forces. Giant step are destroyed completely or in part
are labeled by boxes (The 1:1 resonance is completely destroyed, hence it is
not labeled). The destroyed branches are disordered phases.} 
\end{figure}
 
\begin{figure}
%\narrowtext
\caption{The evolution of $x_j(t)$ for ordered and disordered phases. The
motion on ordered step is quite good travelling wave, while the motion is
irregular and even chaotic in disordered phases.}
\end{figure}
 
\begin{figure}
%\narrowtext
\caption{The v-F characteristics for the approach to golden mean $\delta _G=( 
\sqrt{5}-1)/2$ via the Fibonacci sequence $\delta =\frac 58$, $\frac 8{13}$, 
$\frac{13}{21}$, ... . The disordered phase 1:1 can always be observed.} 
\end{figure}
%\end{multicols}
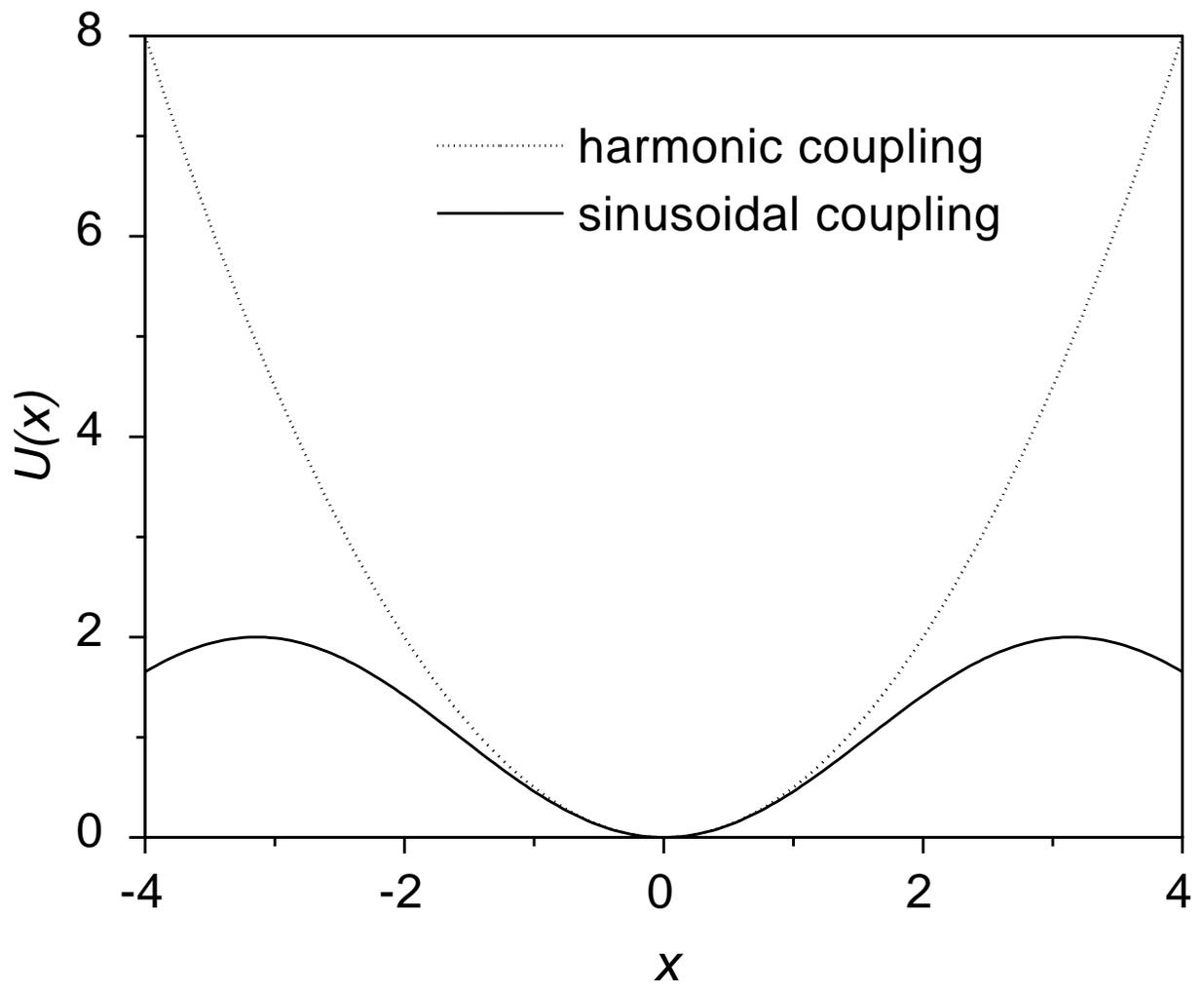
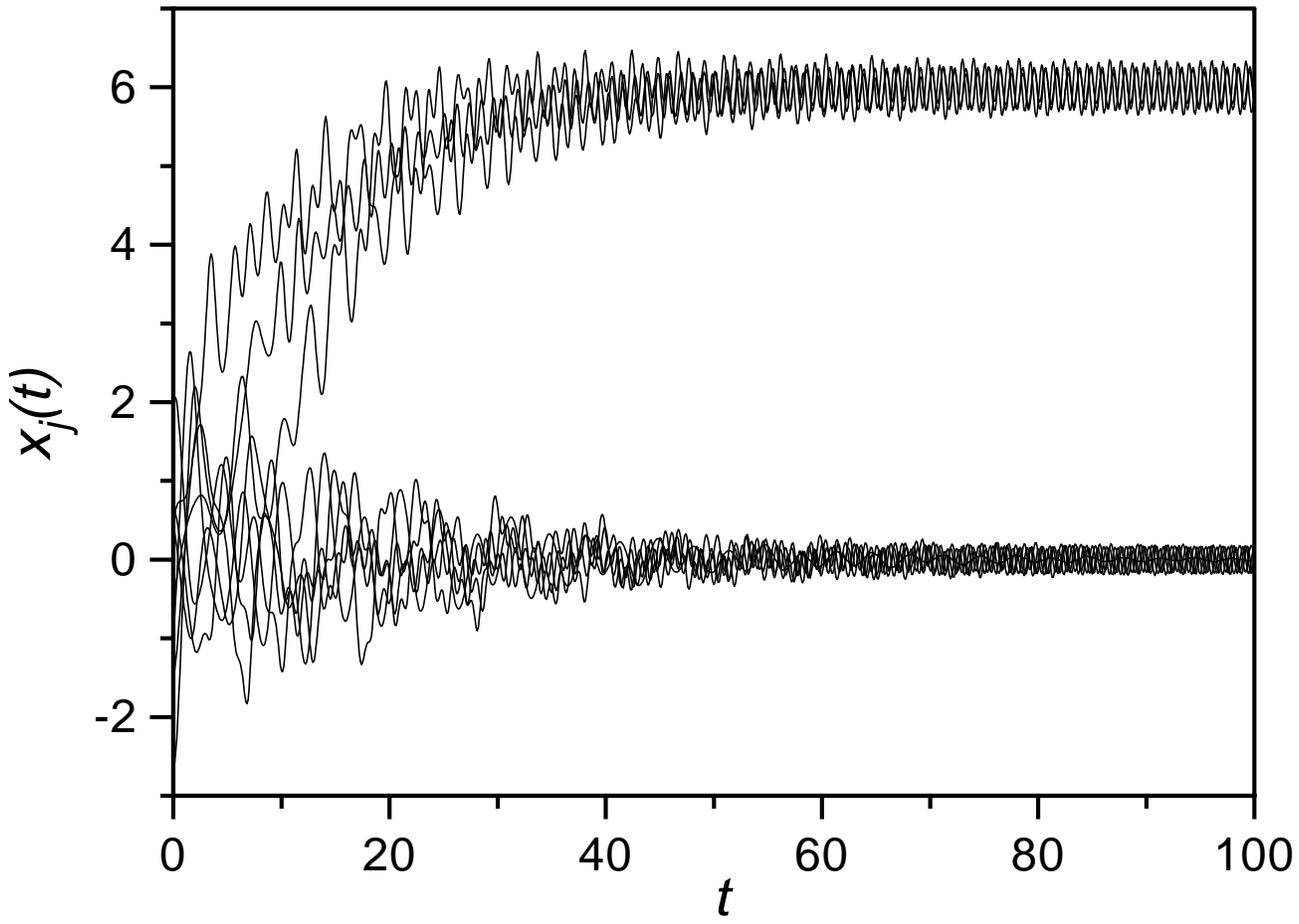
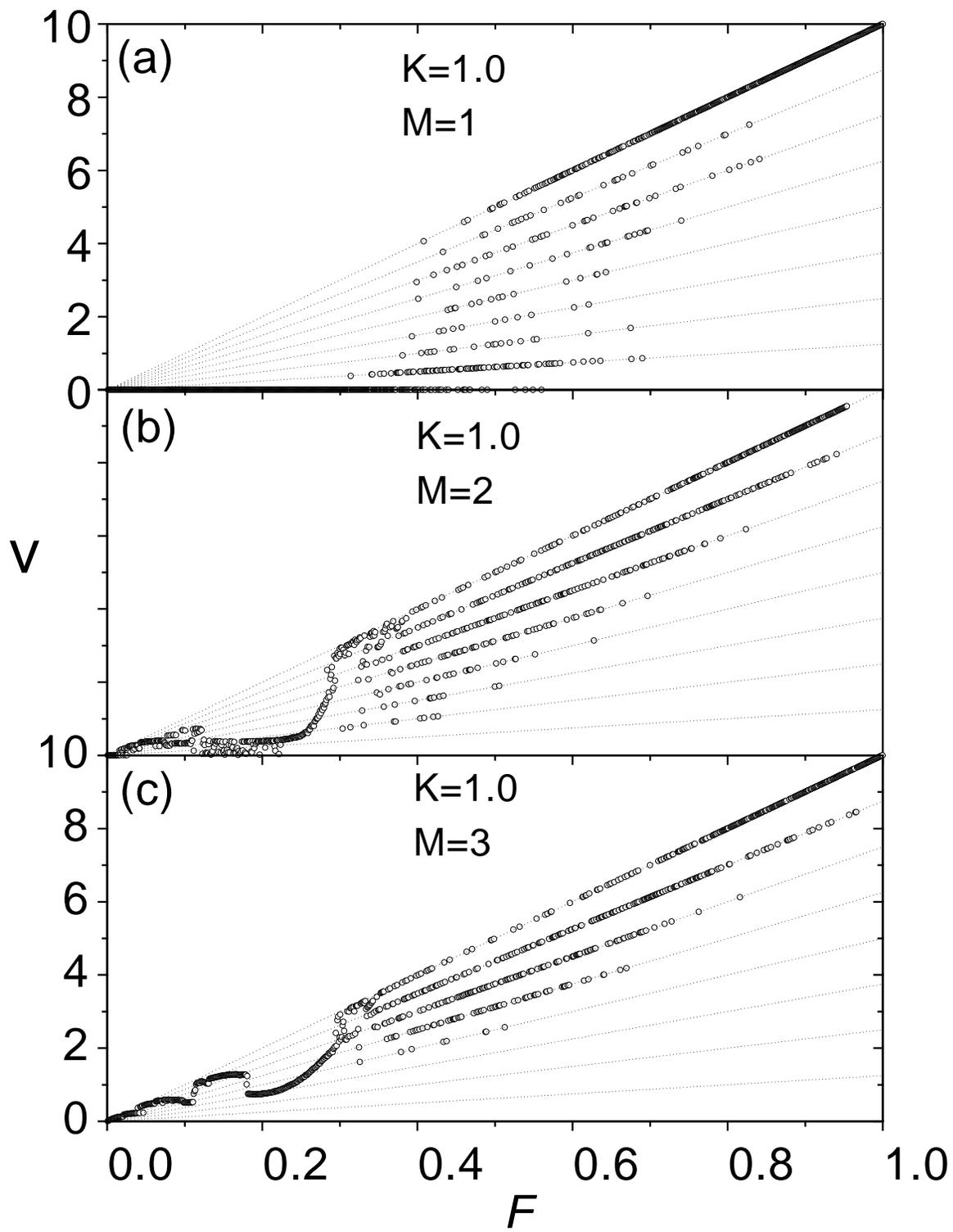
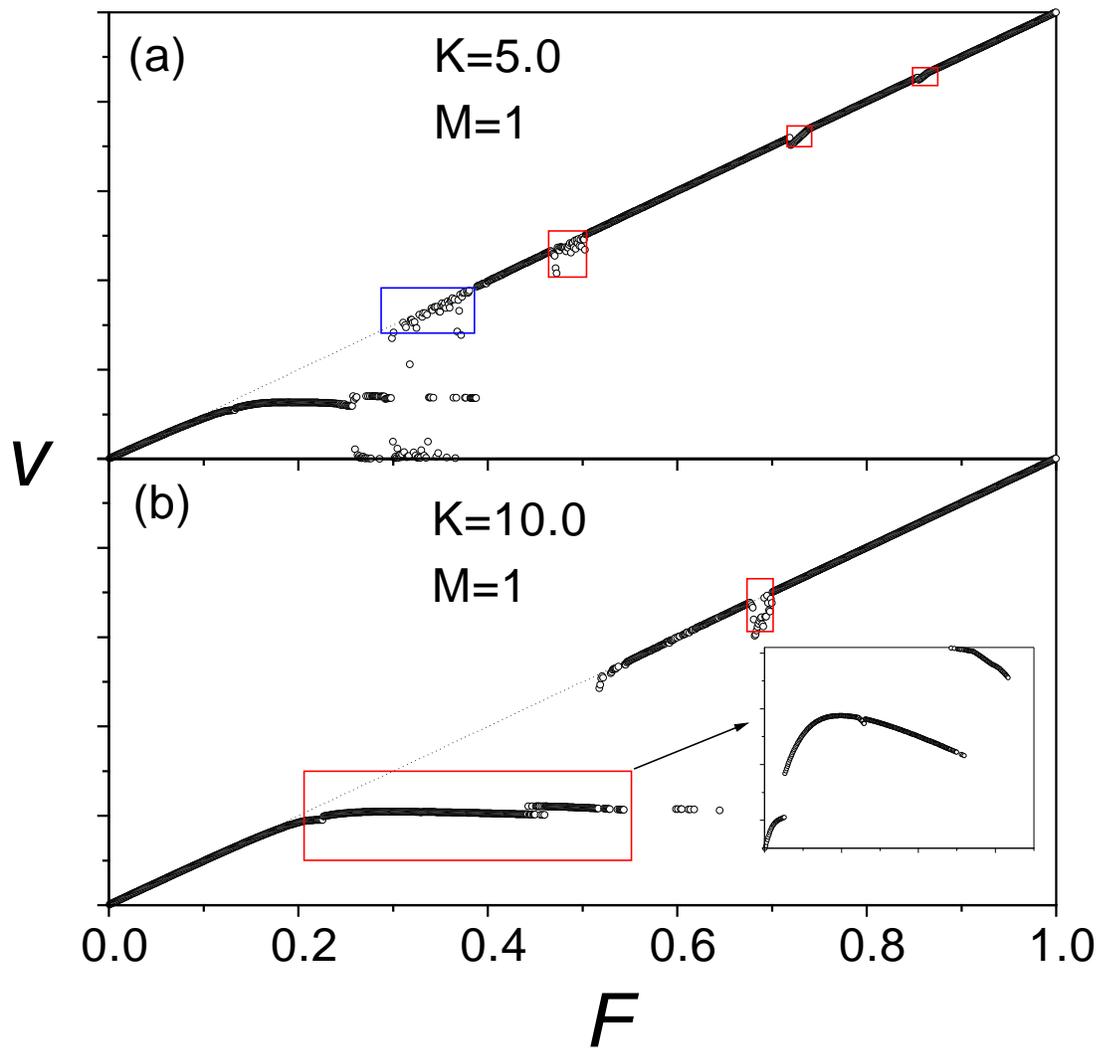
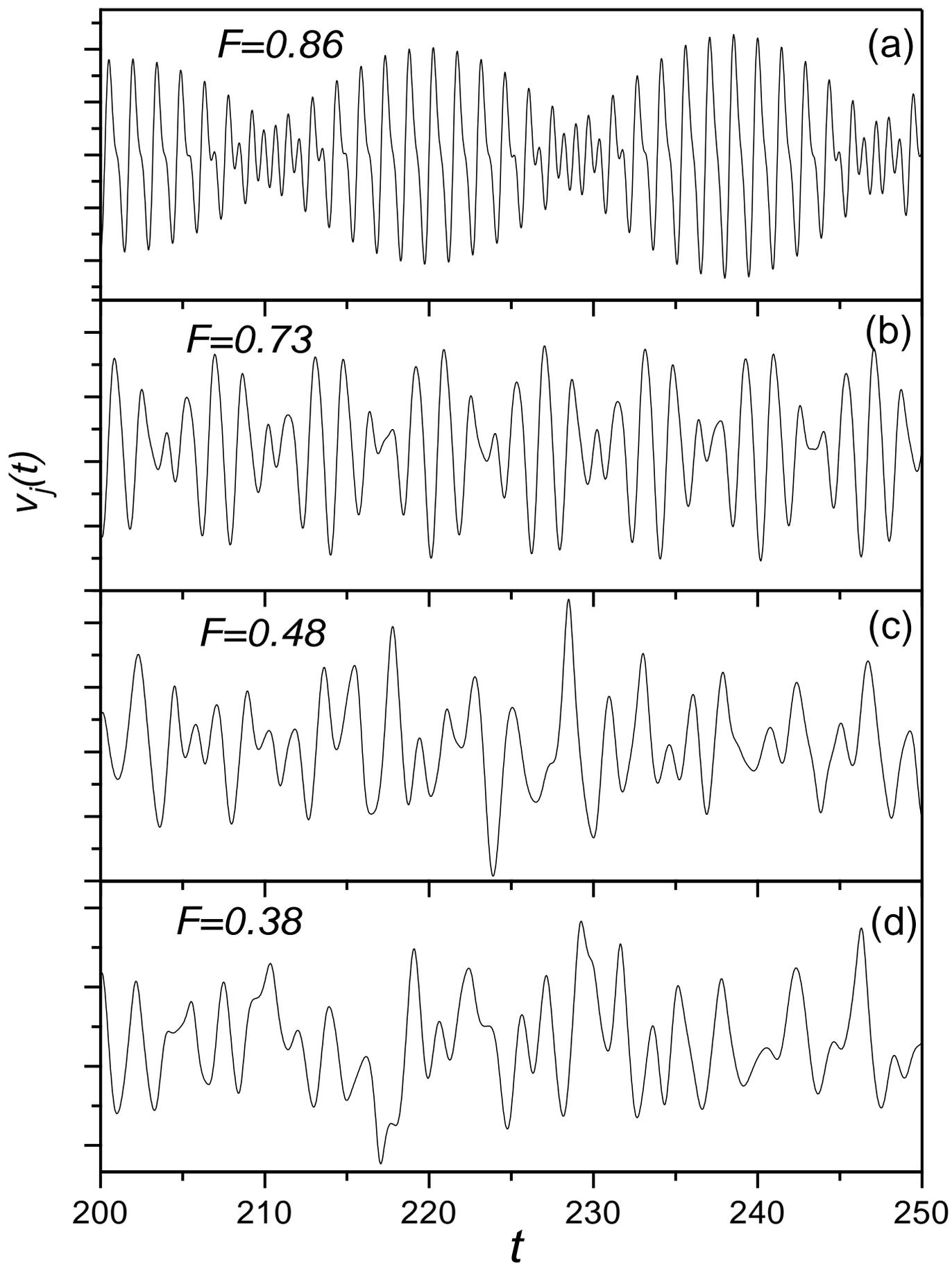
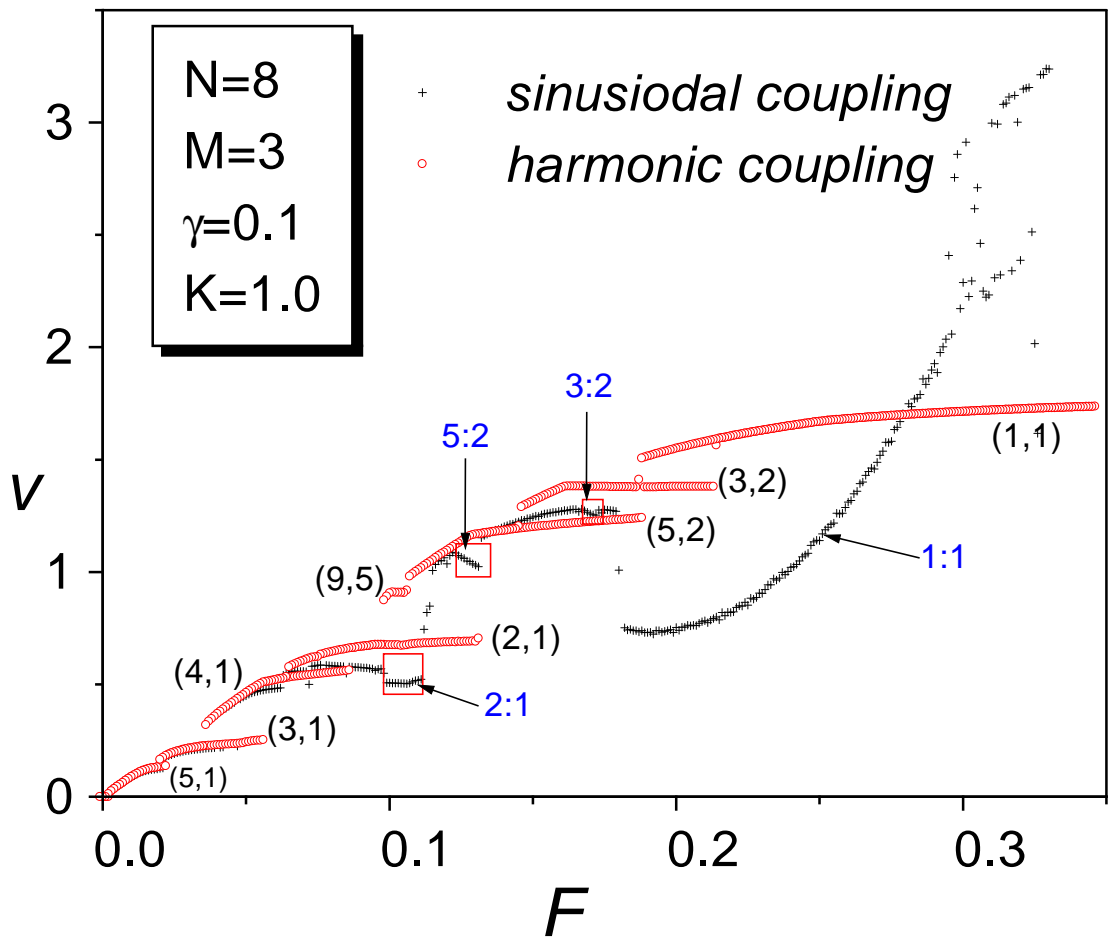
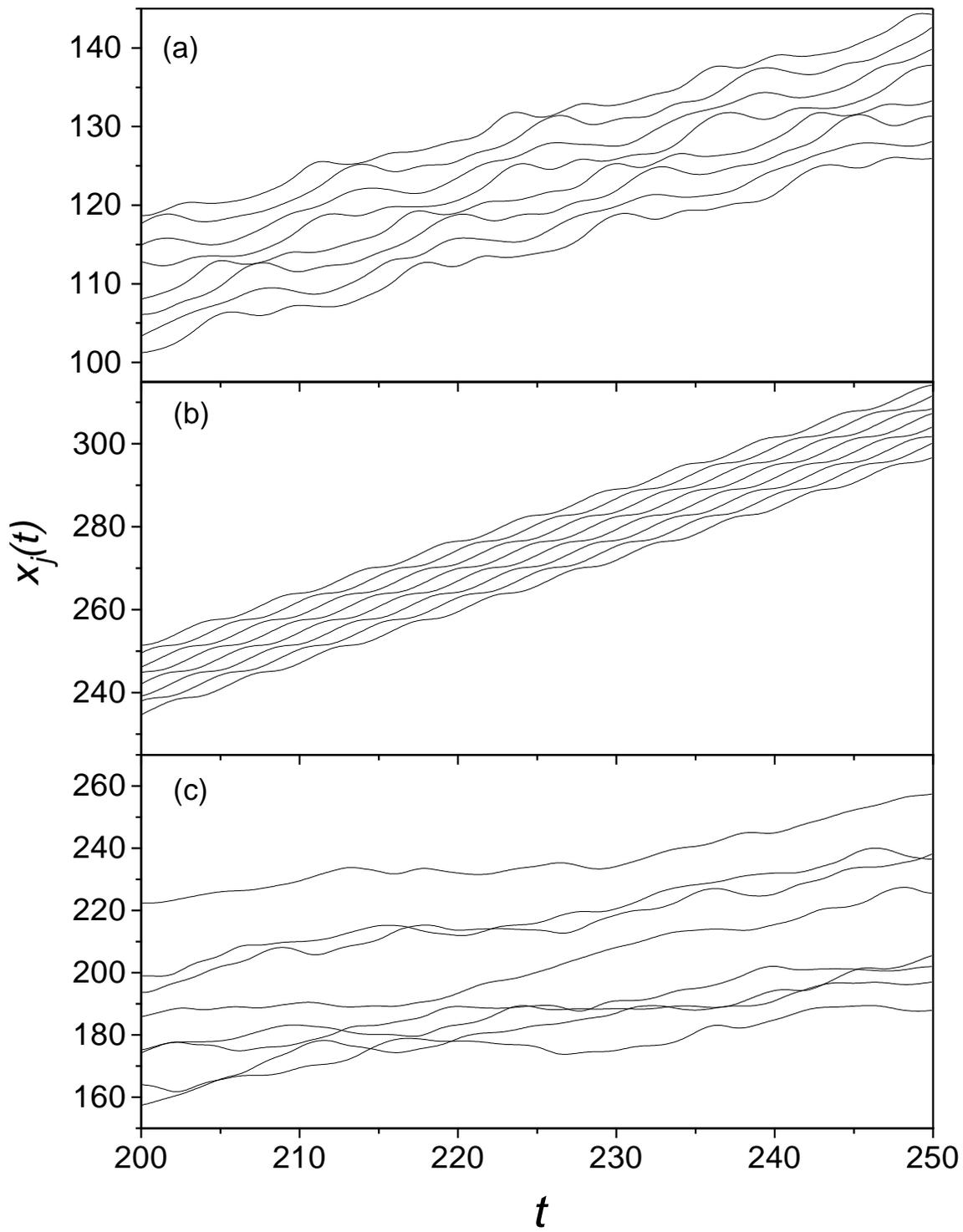
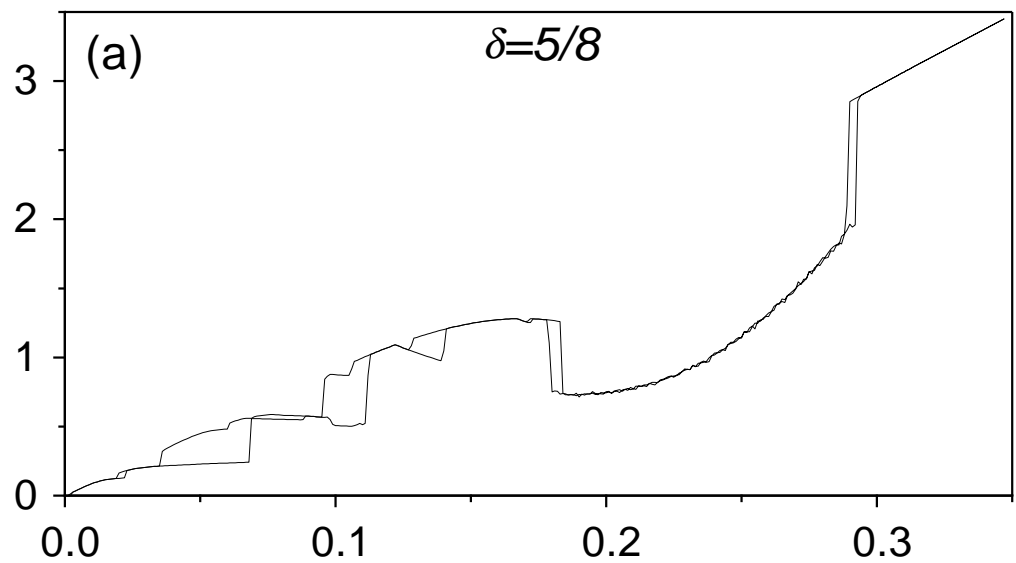
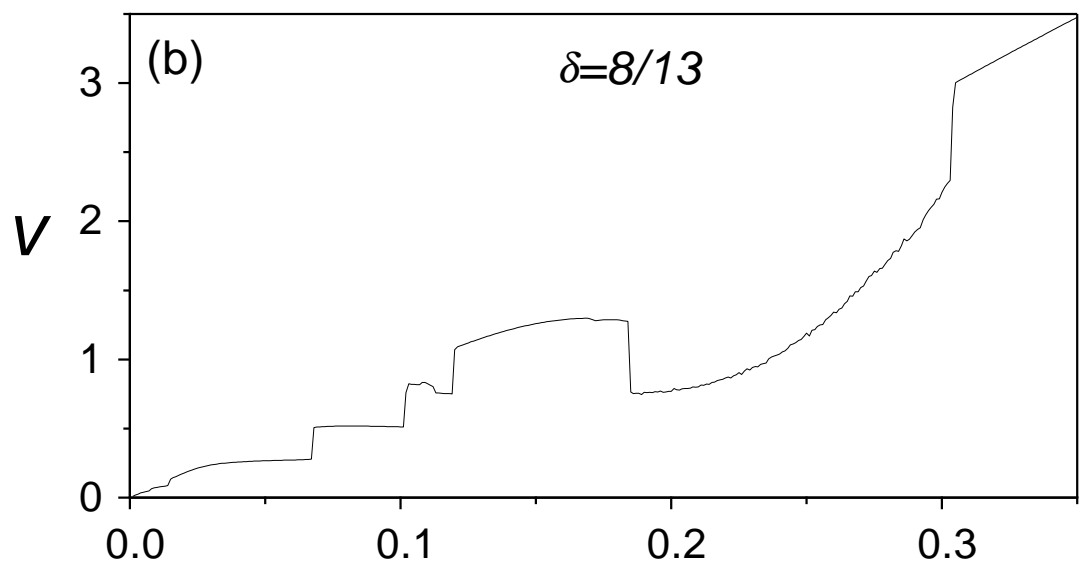
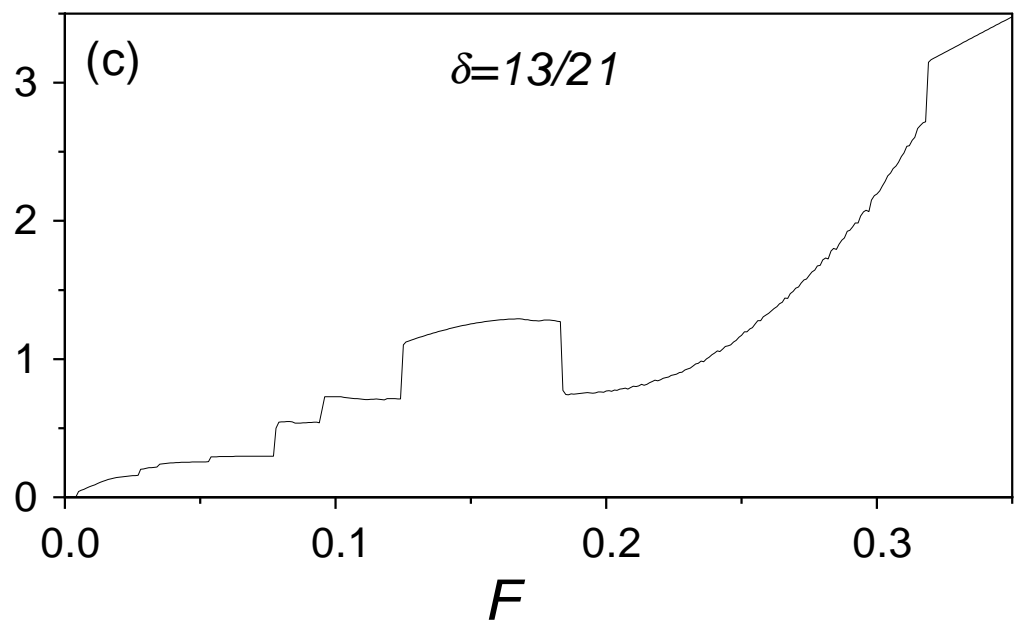
\end{document}